\documentclass[aps, pre, reprint, showpacs,preprintnumbers, a4paper, superscriptaddress]{revtex4-1}
\usepackage[utf8x]{inputenc}
\usepackage[english]{babel}
\usepackage{graphicx}% Include figure files
\usepackage{dcolumn}% Align table columns on decimal point
\usepackage{amsmath}
\usepackage{amssymb}
\usepackage{bm}% bold math
\usepackage{color}        % used for the two-column index
\definecolor{darkred}{rgb}{0.5,0.0,0.0}

\begin{document}
\title{Delay-Induced Depinning of Localized Structures in a spatially inhomogeneous Swift-Hohenberg Model}% Force line breaks with \\
\author{Felix Tabbert}
\email{felix.tabbert@uni-muenster.de}
\affiliation{Institute for Theoretical Physics, University of M\"unster,
Wilhelm-Klemm-Str.\,9, D-48149 M\"unster, Germany}
\author{Christian Schelte}
\affiliation{Institute for Theoretical Physics, University of M\"unster,
Wilhelm-Klemm-Str.\,9, D-48149 M\"unster, Germany}
\author{Mustapha Tlidi}
\affiliation{Faculté des Sciences,\\ Université Libre de Bruxelles,\\ Campus Plaine, C.P. 231,\\ Brussels B-1050, Belgium}
\author{Svetlana V. Gurevich}
%\author{R.~Friedrich}
\affiliation{Institute for Theoretical Physics, University of M\"unster,
Wilhelm-Klemm-Str.\,9, D-48149 M\"unster, Germany}
\affiliation{Center for Nonlinear Science (CeNoS), University of M\"unster,
Corrensstr.\,2,\\ D-48149 M\"unster, Germany\\}
\date{\today}
%\thanks{Corresponding author}
\begin{abstract}
% We report on the dynamics of localized structures  in a Swift-Hohenberg model describing pattern formation in a coherently driven optical resonator with an additional spatial inhomogeneity and subject to time-delayed feedback. The inhomogeneity breaks the translational symmetry of the system by exerting an attracting force on the localized structure. Without time-delayed feedback, the localized structure gets pinned on the center of the inhomogeneity, suppressing the drift bifurcation that has been observed in the same system without inhomogeneities. We provide a linear stability analysis of localized solutions which allows us to predict the different instabilities induced by time-delay feedback. These instabilities are further studied in direct numerical simulations where we report on the formation of two-arm spirals as well as on oscillatory dynamics and depinning dynamics caused by the interplay of an attracting inhomogeneity and destabilizing time-delay. The transition from oscillating to depinning solutions is studied by means of numerical continuation techniques and by using two semi-analytical approaches \ttsveta{TODO}.  

We report on the dynamics of localized structures in an inhomogeneous Swift-Hohenberg model describing pattern formation in the transverse plane of an optical cavity. This real order parameter equation is valid close to the second order critical point associated with bistability. The optical cavity is illuminated by an inhomogeneous
spatial gaussian pumping beam, and subjected to time-delayed feedback. The gaussian injection beam breaks the translational symmetry of the system by exerting an attracting force on the localized structure. We show that the localized structure can be pinned to the center of the inhomogeneity, suppressing the delay-induced drift bifurcation that has been  reported in the particular case where the injection is homogeneous, assumming a continous wave operation. Under an inhomogeneous spatial pumping beam, we perform the stability analysis of localized solutions to identify different instability regimes induced by time-delayed feedback. In particular, we predict the formation of two-arm spirals, as well as oscillating and depinning dynamics caused by the interplay of an attracting inhomogeneity and destabilizing time-delayed feedback. The transition from oscillating to depinning solutions is investigated by means of numerical continuation techniques. Analytically, we use  two approaches based on either an order parameter equation, describing the dynamics of the localized structure in the vicinity of the Hopf bifurcation or an overdamped dynamics of a particle in a potential well generated by the inhomogeneity. In the later approach, the time-delayed feedback acts as a driving force. 
\end{abstract}
%
%
%\keywords{Suggested keywords}
%
\maketitle

\begin{section}{Introduction}

Dissipative localized structures have been theoretically predicted and experimentally observed in various fields of natural science such as biology,
chemistry, ecology, physics, fluid mechanics, and optics (see e.g.,~\cite{akhmediev2008dissipative,Lejeune_tlidi_couteron,meron2004vegetation, tlidi2008vegetation, golomb, thiele2013localized, umbanhowar1996localized,Lioubashevski1999, purwins2010dissipative, mikhailov2006control,liehr2013dissipative, ClercPRE2005,Arecchi19991,Short2008,Lloyd201323}). Localized structures of light in the transverse section of passive and active optical devices are often called cavity solitons. Since the experimental evidence of cavity solitons in semiconductor cavities~\cite{barland2002cavity,Ackemann2009323,Kuszelewicz2010,Barbay2011,Averlant:14}, they have attracted growing interest in the nonlinear optics community due to potential applications for e.g., all-optical delay lines or logic gates~\cite{Jacobo2012,PedaciAPL2008}.  Recently, much attention was paid to the investigation of the influence of delayed optical feedback on the stability properties of these structures \cite{tlidi2009, gurevich2013instabilities, panajotov2013optical, panajotov2014chaotic}. Delayed feedback control is a well-established technique that has been applied to various nonlinear  systems (see, e.g.,~\cite{Friedel1998, BestehornPRE2004,schoell2005control,scholl2008handbook, GreenSIAM2009,PaulauPRE2008, tlidi2013delayed, GurevichRD, gurevich2014time, KraftGurevich2016raey} and references thereafter). In particular, it was theoretically demonstrated that a simple time-delayed feedback loop provides a robust and controllable mechanism responsible for the motion as well as for complex oscillatory dynamics of localized structures and localized patterns (see e.g., \cite{tlidi2009,gurevich2013instabilities,panajotov2014chaotic,gurevich2014time,PimenovPRA13,PuzyrevPRA16,PanajotovPRA16}). Especially for the case of the delay-induced motion in a homogeneous system it was shown, that the neutrally stable modes, so-called Goldstone modes, that exist due to the translational invariance of the system under consideration are destabilized by time-delayed feedback, leading to a drift of the localized structure. Since the existence of Goldstone modes only depends on the symmetries of the system in question and its solutions, this behavior can be observed in any system with continuous symmetries possessing localized solutions. However, for a more realistic description of any experimental setup, it is often necessary to take into account spatial inhomogeneities that break the translational symmetry of the system and thus change the dynamics induced by delay. Recently, the competition between a drifting localized structure and spatial inhomogeneities has been studied experimentally in \cite{Caboche2009} and theoretically in a Swift-Hohenberg model
\cite{parra2013dissipative,parra2016competition}, although in the latter case the drift of the localized structure has been introduced by simply adding an advection term to the Swift-Hohenberg equation. 
% 
% Recently, the competition between a drifting localized structure and spatial inhomogeneities in a Swift-Hohenberg model has been studied in \cite{parra2013dissipative,parra2016competition}, where the drift of the localized structure has been introduced by adding an advection term to the classical Swift-Hohenberg equation. 
%   
% The influence of inhomogeneities on drifting localized structures has been 
% studied experimentally in \cite{Caboche2009} and theoretically in 
% \cite{parra2013dissipative}, although in the latter case the drift has been 
% introduced by simply adding an advection term to the Swift-Hohenberg equation and not by considering time-delayed feedback.
  
In this paper, we investigate the competition between unstable translational modes due to delay and spatial inhomogeneities. For this purpose, we consider a passive cavity filled with a two-level medium and driven by a coherent radiation beam and focus on the regime of nascent optical bistability where the spatiotemporal dynamics is described by the Swift-Hohenberg
equation with time-delayed feedback. We show that the inclusion of spatial inhomogeneities  strongly alters the delay-induced dynamics of localized structures.
In particular, two different dynamical solutions resulting from unstable translational eigenmodes are discussed analytically and numerically. For small or moderate values of the delay strength, the localized structure oscillates around the inhomogeneity,  whereas for larger delay strengths the structure depins from the inhomogeneity and drifts freely. Continuation techniques are used to further examine the transition between the two solutions.
% The study of delay-induced drift in this context is in our opinion necessary for future applications in experimental setups as well as instructive since the addition of time-delayed feedback changes the properties of the translational modes which, in combination with the spatial inhomogeneities, allows unstable translational eigenfunctions with complex eigenvalues, i.e. oscillations.

The paper is organized as follows: In section~\ref{sec:Model} the inhomogeneous Swift-Hohenberg model with time-delayed feedback is introduced. In the next section \ref{sec:linstab} the linear stability analysis of localized solutions both without and with spatial inhomogeneities is discussed. In the following section \ref{sec:DNS} results from direct numerical simulations as well as results obtained from numerical continuation techniques are presented. In section \ref{sec:analytic}, two semi-analytic approaches are presented, which are able to account for the transitions from a stable localized structure to an oscillating solution and to a drifting solution respectively. We conclude in section VI.

\end{section}
\begin{section}{The Model}\label{sec:Model}

In this paper we study the effects of spatial inhomogeneities on the dynamics of localized structures in a Swift-Hohenberg model with time-delayed feedback. The model describes the time evolution of the electrical field envelope in the transverse plane of a passive cavity. This simple real order-parameter equation is valid only close to a second-order critical point associated with the optical bistability. The cavity is driven by an injection beam and is subjected to time-delayed feedback. This model was extensively studied in \cite{tlidi1994localized,tlidi2009,tlidi2010spontaneous,gurevich2013instabilities}. The Swift-Hohenberg model with time-delayed feedback and a homogeneous injection beam can be written as
% \begin{align}
% \partial_t \textbf{q}(\textbf{x},t)=-a_1\Delta 
% \textbf{q}(\textbf{x},t)-a_2\Delta^2 
% \textbf{q}(\textbf{x},t)+Y_0+C\textbf{q}(\textbf{x},t)-\textbf{q}(\textbf{x},
% t)^3+\alpha(\textbf{q}(\textbf{x},t)-\textbf{q}(\textbf{x},t-\tau))\label{DSH},
% \end{align}
% \begin{align}
% \partial_t q(\textbf{x},t)&=-a_1\Delta 
% q(\textbf{x},t)-a_2\Delta^2 
% q(\textbf{x},t)+f(q),\\
% f(q)&=Y_0+Cq(\textbf{x},t)-q(\textbf{x},
% t)^3+\alpha(q(\textbf{x},t)-q(\textbf{x},t-\tau))\label{DSH},
% \end{align}
\begin{align}
\partial_t q_t=\left(-a_1\Delta 
-a_2\Delta^2+C 
\right)q_t+Y_0-q_t^3+\alpha(q_t-q_{t-\tau})\label{DSH},
\end{align}
 where the state variable $q_t=q(\textbf{x},t)$ and the scalar quantity $Y_0$ represent the deviation of the internal and injected field from their values at the critical point, respectively. the Laplacian $\Delta=\partial_x^2+\partial_y^2$ acts on the transverse plane $\textbf{x}=(x,y)$.  $C$ represents the deviation of the cooperativity parameter from its critical value, $a_1$, $a_2 > 0$ are positive constants obtained by rescaling during the derivation of Eq. \eqref{DSH} \cite{tlidi2010spontaneous}. 

The time-delayed feedback can be implemented in an experimental setup by adding an external mirror at some distance $L$ from the cavity. The light will undergo an excursion in the external cavity and  will be reinjected into the cavity. The delay time is $\tau=2L/c$ where $c$ is the speed of the light. The delay strength $\alpha$  is proportional
to the reflectivity of the external mirror  and inversely proportional to the Fabry-Perot round trip time. The phase of the feedback is fixed to $\pi$. For a more detailed description of the  derivation of Eq. \eqref{DSH} as well as of the physical meaning of the parameters we refer the reader to \cite{tlidi2010spontaneous}.

Without the delayed feedback i.e., $\alpha=0$, equation 
(\ref{DSH}) posseses a Lyapunov functional that decreases monotonically in the course of time
\cite{cross1993pattern} and has been studied in various fields of
nonlinear science~\cite{Hilali1996263, Lefever2009194, TlidiGSH2003, Stoop2015337, BurkeKnoblochChaos2007, BurkePRE2006, Andrei11}. However, in the presence of time-delayed feedback the system looses its gradient structure. 

In translationally invariant systems that posses a Lyapunov functional, a drift bifurcation is shown to be the first instability induced when adding time-delayed feedback due to a destabilization of Goldstone modes associated with the continous symmetries of the system \cite{friedrich1993higher, KraftGurevich2016raey}. Even in systems without a gradient structure a drift bifurcation is the first occurring instability in a wide parameter regime \cite{GurevichRD}.  However, physically realistic systems are never completely invariant under translation, due to boundaries of the system and due to spatial inhomogeneities. In the following, we are going to concentrate on the effects of spatial inhomogeneities  on the space time evolution of the intracavity field subjected to the time-delayed feedback. The spatial inhomogeneity originates from the fact that the injected beam is not uniform in the transverse plane. We consider a gaussian injection beam instead of a continuous wave operation
% The influence of inhomogeneities on drifting localized structures has been 
% studied experimentally in \cite{Caboche2009} and theoretically in 
% \cite{parra2013dissipative}, although in the latter case the drift has been 
% introduced by simply adding an advection term to the Swift-Hohenberg equation and not by considering time-delayed feedback. 
\begin{align}\label{eq:Inhom}
Y(\textbf x)=Y_0+A\text{e}^{\frac{\textbf  - x^2}{B}},
\end{align}
where $A$ is the amplitude of the inhomogeneity and $B$ is the width of the gaussian. The introduction of the inhomogeneous injection field breaks the translational symmetry of the 
system, i.e. the parameter $Y=Y(x,y)$ in Eq. \eqref{DSH} now depends explicitly on the spatial coordinates.

The inhomogeneity alters the stationary solutions of the system \cite{parra2013dissipative}. In particular, the homogeneous solution of Eq. \eqref{DSH} without time-delayed feedback becomes deformed, showing a low bump at the center of the inhomogeneity. Another stationary solution consists of a localized structure pinned at the inhomogeneity. The height and width of this localized structure depend on the strength of the inhomogeneity, i.e. the structure grows with increasing amplitude $A$ or increasing width  $B$. It has been experimentally and analytically shown that the inhomogeneity of the pump allows to stabilize localized structures resulting from fronts connecting two homogeneous steady states \cite{Odent2014}. Apart from the localized structure positioned directly on the inhomogeneity, there are several other stationary localized solutions. If the strucutre initially is positioned in the vicinity of the inhomogeneity it gets either pulled to its center or it gets repelled if the initial distance to the center is too large, creating a stable solution next to the inhomogeneity. In the following sections we will focus on the impact of time-delayed feedback on a localized structure sitting in the center of the inhomogeneity ${Y}(\mathbf{x})$, given by Eq.~\eqref{eq:Inhom}.
\end{section}

\begin{section}{Linear Stability Analysis}\label{sec:linstab}
As a first approach to analyze the destabilization of a localized structure by time-delayed feedback, we perform a linear stability analysis of the system. Linearizing Eq. \eqref{DSH} for $\alpha=0$ around the stationary solution $ q_0(\textbf x)$ yields a linear operator $\mathfrak L[q_0(\textbf x)]$. By solving the resulting linear eigenvalue problem $\mathfrak L[q_0(\textbf x)] \varphi_k(\textbf x)=\mu_k  \varphi_k(\textbf x)$ numerically, one obtains the eigenvalues $\mu_k$ of the undelayed system as well as the corresponding eigenfunctions $\varphi_k(\textbf x)$. In the following we first analyze the spectrum of a localized structure in the homogeneous case both without and with time-delayed feedback and then discuss the changes in the discrete spectrum of the problem in question induced by the introduction of a spatial inhomogeneity.

\begin{subsection}{Linear Stability Analysis without inhomogeneities}

In the case of a single stationary localized structure $q_0(\textbf{x})$ (see Fig. \ref{fig:EW1s}~(a)), the spectrum of the linearized Swift-Hohenberg operator consists of a discrete part close to zero with corresponding localized eigenfunctions and a well separated continuous part \cite{gurevich2013instabilities}. Without time-delay, the solution is stable, i.e. all real eigenvalues $\mu\le0$. Neglecting inhomogeneities, the discrete part of the spectrum in 2D consists of two eigenvalues $\mu=0$ with eigenfunctions that correspond to an infinitesimal translation of the structure in two spatial directions. These so-called Goldstone modes are neutrally stable due to the translational invariance of Eq. \eqref{DSH} \cite{friedrich1993higher}. One of these two modes is depicted in Fig. \ref{fig:EW1s} (b). The discrete spectrum corresponding to $\mu<0$ consists of one eigenfunction that would lead to a growth or shrinkage of the structure (Fig. ~\ref{fig:EW1s}(c)) and two eigenfunctions that 
correspond to a deformation of the structure in different spatial directions (Fig. ~\ref{fig:EW1s}(d))~ \cite{gurevich2013instabilities}.

Applying time-delayed feedback, i.e. $\alpha\neq0$, $\tau\neq0$, does neither change the stationary solutions of the systems nor the eigenfunctions, due to the special form of the delay term in equation \eqref{DSH}, which is often refered to as Pyragas control \cite{pyragas1992}. However, the time-delayed feedback changes the eigenvalues of each eigenfunction, i.e., it may change the stability of the stationary solutions. The new eigenvalues $\lambda_{k,m}$ with time-delay can be calculated as \cite{gurevich2013instabilities}:
\begin{align}
\lambda_{k,m}=\mu_k+\alpha+\frac{1}{\tau}W_m\left[-\alpha \tau \cdot 
e^{-\tau(\mu_k+\alpha)}\right],~~~~~~~~m\in\mathbb Z \label{DEW},
\end{align}
where $W_m$ is the $m$-th branch of the Lambert W function \cite{LambertW}. As shown in Eq. \eqref{DEW}, the addition of time-delayed feedback creates an infinite amount of complex eigenvalues $\lambda_{k,m}$ for each real-valued eigenvalue of the undelayed 
system $\mu_k$, due to the multivalued character of the Lambert W functions 
$W_m$. 

Since we are interested in the destabilization of the discrete spectrum of the localized solution ${q_0}$ our main interest lies in the eigenvalues $\lambda_{k,0}$, corresponding to the main branch of the Lambert W function, because these are the eigenvalues with the highest real parts, i.e., the first eigenvalues to become unstable.

As mentioned above, the first eigenfunctions which become unstable with increasing time-delayed feedback parameters are the Goldstone modes leading to the above mentioned drift of the localized structure. Without inhomogeneities, the stability threshold $\alpha\tau=1$ of these eigenfunctions can be calculated analytically \cite{tlidi2009,gurevich2013instabilities}. By increasing the time-delayed feedback further, one can induce instabilities of other localized eigenfunctions or even destabilize the homogeneous background of the localized structure, thus inducing traveling waves or homogeneous oscillations \cite{gurevich2013instabilities, KraftGurevich2016raey}.

Note that instead of solving Eq. \eqref{DEW} for different values of $\alpha$ and $\tau$ to calculate the stability threshold of a given eigenfunction corresponding to an eigenvalue $\mu_k$, one can use the following expression to determine the critical delay-time $\tau_c$ that induces a change of stability \cite{gurevich2014time}:
\begin{align}
 \tau_c=\frac{\pm\arccos\left(1+\frac{\text{Re}(\mu_k)}{\alpha}\right)+2\pi n}{\text{Im}(\mu_k)\pm\alpha\sqrt{1-\left(1+\frac{\text{Re}(\mu_k)}{\alpha}\right)^2}},~~~~~~~~n\in\mathbb N.\label{taukritisch}
\end{align}
Note that, besides $\alpha$, the stability threshold in general also depends on both the real and the imaginary part of the eigenvalue $\mu_k$.

\begin{figure}[!ht]
 \includegraphics[width=0.45\textwidth]{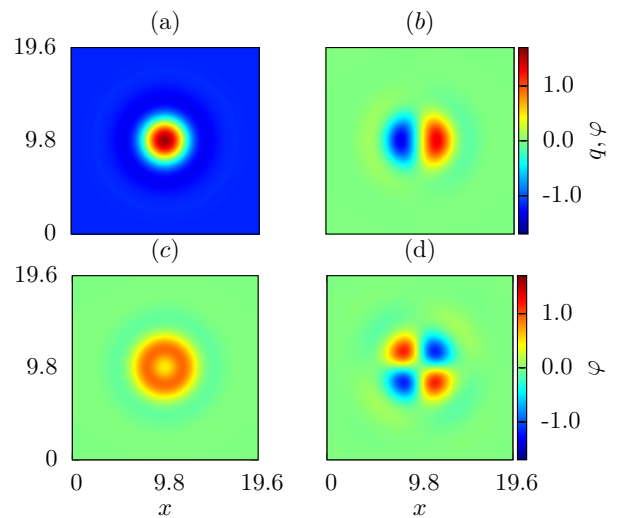}
\vspace{0.3cm}
 \caption{(a) Stable localized solution $q_0 (\textbf x)$ obtained by 2D numerical integration of Eq.~\eqref{DSH} as well as three localized eigenfunctions $\varphi_k (\textbf x)$ as solutions of the linear eigenvalue problem for $\alpha=0$: (b) translational mode; (c) growth mode; (d) deformation mode. Other parameters are: $L_x=L_y=19.6$, $a_1=2.0$, $a_2=\frac{4}{3}$, $Y_0=-0.4$, $C=1.0$, $A=0$, $B=0$.}
 \label{fig:EW1s}
\end{figure}
\end{subsection}

\begin{subsection}{Linear Stability Analysis with inhomogeneities}
Considering now a localized solution positioned in the center of the inhomogeneity in the full inhomogeneous system, one can proceed in the same way as in the homogeneous case, i.e. first performing the linear stability analysis without delay and then calculating the eigenvalues with delay using Eq. \eqref{DEW}.
Although the localized solution as well as the eigenfunctions change slightly in the presence of the inhomogeneity compared to the homogeneous case, one can still clearly identify two translational modes, one growth mode and two deformation modes as they are depicted in Fig. \ref{fig:EW1s} (b-d). However, even without time-delayed feedback, the eigenvalues corresponding to each eigenfunction change, compared to the homogeneous case. 

The eigenvalues of the different localized eigenfunctions as a function of the amplitude $A$ of the inhomogeneity are shown in Fig. \ref{mudefekt}. In the absence of an inhomogeneity ($A=0$), the translational eigenfunction is neutrally stable, i.e., corresponds to $\mu=0$. By increasing $A$ the eigenvalue gets lowered and the structure gets pinned on the inhomogeneity. As can be seen in Fig. \ref{mudefekt}, the order of the eigenvalues changes with increasing amplitude $A$ of the inhomogeneity, leading to two different regimes: For small amplitudes the drift-inducing translational modes still posses the
highest eigenvalue $\mu$. For larger amplitudes, however, the deformation-inducing modes become the modes with the highest eigenvalue. Note that changing the width $B$ of the inhomogeneity instead of its amplitude basically reproduces the same behavior of the eigenvalues $\mu_k$.

\begin{figure}[h]
 \centering
\includegraphics[width=0.5\textwidth]{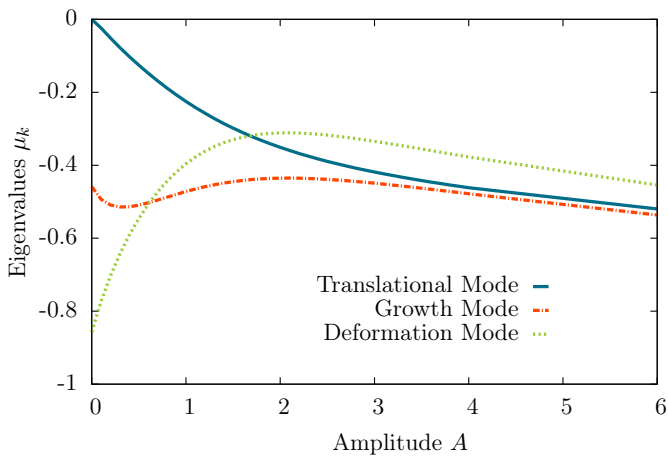}
 \caption{Eigenvalues $\mu_k$ of Eq.~\eqref{DSH} for $\alpha=0$ corresponding to the drift-inducing modes (blue solid
line), the growth-inducing mode (red dash-dotted line) and the deformation-inducing modes 
(green dotted line) for different amplitudes $A$ of the inhomogeneity $Y$.}
 \label{mudefekt}
\end{figure}

Adding time-delayed feedback in the inhomogeneous case can be treated in the same way as in the homogeneous case, i.e. the stability thresholds can be calculated using Eq. \eqref{taukritisch}.
 Considering the destabilization of the translational modes, one should note, that even for only real-valued $\mu_k$, the corresponding eigenvalues $\lambda_{k,0}$ are generally complex, thus allowing for oscillatory dynamics. In case of the translational mode, the two highest eigenvalues $\lambda_{0,0}$ and $\lambda_{0,-1}$ stay real, if the original eigenvalue without delay is $\mu=0$, i.e. in the homogeneous case, whereas they become complex, if $\mu\neq 0$. That is, by adding a inhomogeneity to the system, one changes the dynamics induced by an unstable translational mode drastically, allowing oscillatory behavior.
\end{subsection}

\end{section}

\begin{section}{Direct Numerical Simulations}\label{sec:DNS}
Once one has calculated the stability thresholds of different localized eigenfunctions using Eq. \eqref{taukritisch}, it is necessary to determine how unstable eigenfunctions affect the dynamics of the system and which eigenfunctions govern the dynamics in regions of multiple instabilities. Therefore we perform direct numerical simulations of Eq. \eqref{DSH} for different values of $\alpha$ and $A$ using a semi-implicit Euler timestepping and a pseudospectral method on a periodic domain to calculate the spatial derivatives of $q(\textbf{x},t)$. 

Fixing the delay time at $\tau=1$ and the width of the inhomogeneity at $B=4$, and varying the delay strength $\alpha$ and the amplitude $A$ of the inhomogeneity, different dynamical solutions can be observed. In particular, starting with a small amplitude $A=0.2$ one can identify three dynamical regimes. For delay strengths $\alpha<\alpha_\text{crit}=1.0219$ the localized structure is still stable. For values $\alpha\ge\alpha_\text{crit}$ the translational mode becomes unstable inducing a movement of the localized structure. However, the inhomogeneity still has an attracting effect on the localized structure, i.e. the inhomogeneity pulls the structure back, leading to an oscillatory motion of the localized structure around the defect as can be seen in Fig. \ref{fig:1d1szeit} on the left. The possibility of such a periodic behavior is evident, considering that the highest eigenvalue $\lambda_0$ is now complex due to the inhomogeneity. In fact, the frequency $\omega$ of the oscillatory behavior at the 
bifurcation point coincides with the complex part of the eigenvalue $\text{Im}(\lambda_0)$ (cf. Fig. \ref{fig:omega}). Note, that the onset of the instability at $\alpha_\text{crit}$ can be obeserved both in the linear stability and in the direct numerical simulations.

\begin{figure}[!ht]
\centering
\includegraphics[width=0.5\textwidth]{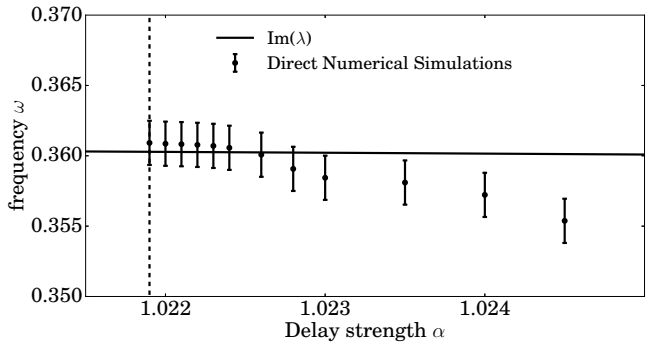}
 \caption{Frequency $\omega$ of the oscillations of the localized structure around the inhomogeneity obtained from direct numerics in one dimension (dotted line) and the imaginary part of the eigenvalue corresponding to the unstable translational mode (solid line). The dotted vertical line marks the onset of the oscillations, i.e. the first bifurcation point.}
 \label{fig:omega}
\end{figure} 

With increasing delay strength $\alpha$, the effect of the unstable translational mode increases too, resulting in a larger amplitude of the oscillations. Eventually the delayed feedback leads to a destabilization of the periodic solution, i.e. the localized structure gets depinned from the inhomogeneity and starts to drift freely (see Fig. \ref{fig:1d1szeit}, on the right). For the sake of simplicity, the simulations in Fig. \ref{fig:1d1szeit} are performed in one spatial dimension. However, one can observe similar dynamics in two spatial dimensions, where only the value of $\alpha_\text{crit}$ changes slightly. In the oscillatory regime, phase-independent oscillations in two spatial directions lead to a wiggling motion of the structure around the defect. For larger values of $\alpha$, the localized structure depins, however, the direction of the occurring drift is arbitrary.  
\begin{figure}[!ht]
 \centering
\includegraphics[width=0.54\textwidth]{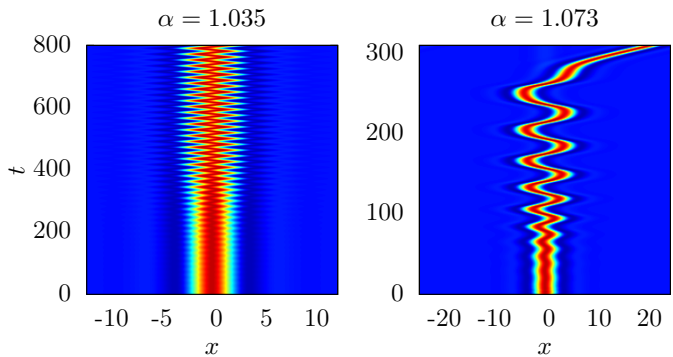}
 \caption{Direct numerical simulations in one dimension with a fixed delay-time $\tau=1$.  Left panel: A localized solution oscillates around the inhomogeneity for  $\alpha=1.035$. Right panel: A localized structure gets depinned from the inhomogeneity and starts to drift freely for $\alpha=1.073$. The amplitude of the inhomogeneity is fixed to $A=0.2$. Other parameters are: $a_1=2.0$, $a_2=\frac{4}{3}$, $Y_0=-0.4$, $C=1.0$, $B=4.0$.}
 \label{fig:1d1szeit}
\end{figure}

%\ttsveta{XXXX}
In order to investigate the transition from a bound oscillatory movement to a free drift in detail, we used path continuation techniques provided by the Matlab package DDE-BIFTOOL \cite{engelborghs2002numerical} for delay differential equations. To this aim the behavior of the system in one spatial dimension $x$ has been investigated. Considering that DDE-BIFTOOL is designed to continuate delay differential equations, we decompose Eq. \eqref{DSH} into a set of coupled delay differential equations by replacing the spatial derivatives with finite differences.

Continuation of the period time $T$ of the periodic solution in $\alpha$ shows, that a stable limit cycle evolves at $\alpha_\text{crit}$ and looses its stability in a bifurcation at a critical value $\alpha_\text{{crit}2}>\alpha_\text{crit}$ (see Fig. \ref{fig:Bifdiag}). At this bifurcation point $\alpha_\text{{crit}2}$, the localized structure gets depinned from the inhomogeneity in the direct numerical simulations. Looking at the Floquet multipliers one can identify the bifurcation as a saddle-node bifurcation of a limit cycle, where the stable limit cycle (black line) merges with an unstable one (red dotted line) and anihilates.
\begin{figure}[!ht]
 \centering \includegraphics[width=0.45\textwidth]{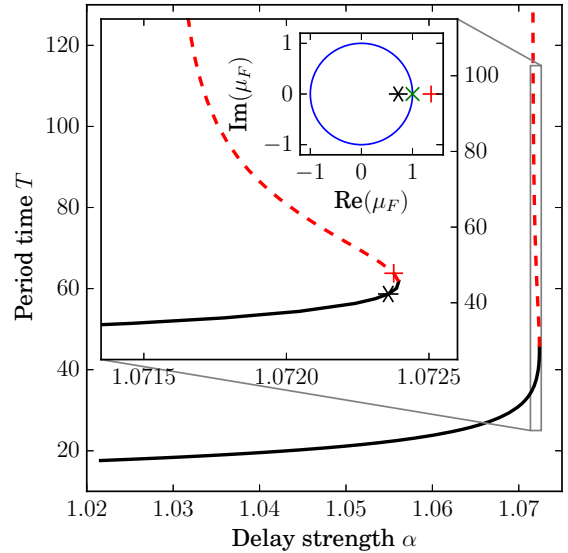}
\vspace{0.3cm}
 \caption{Period time $T$ of the oscillatory solution depending on the delay strength $\alpha$ obtained by numerical continuation using DDE-BIFTOOL. The zoomed in version shows the parameter region in the vicinity of the bifurcation point, where the saddle-node bifurcation of limit cycles sets in. The small box at the top shows the Floquet multipliers of the stable (black) and the unstable (red) periodic branch at the marked positions.}
 \label{fig:Bifdiag}
\end{figure} 

As shown in the previous section, the order of the eigenvalues $\mu_k$ without delay changes with an increasing amplitude $A$ of the inhomogeneity. In particular, for $A=2$ in two dimensions, the deformation modes are the ones with the largest eigenvalue, i.e., the first modes to be destabilized by time-delayed feedback. A destabilization of these modes leads to a deformation of the localized structure as shown in Fig.~\ref{fig:EW1s}. Due to this deformation, the localized structure looses its rotational symmetry, which in combination with the time-delayed feedback leads to a rotation of the spreading spiral structure (see Fig. \ref{fig:spirale}). 
\begin{figure}[!ht]
 \centering
 \includegraphics[width=0.5\textwidth]{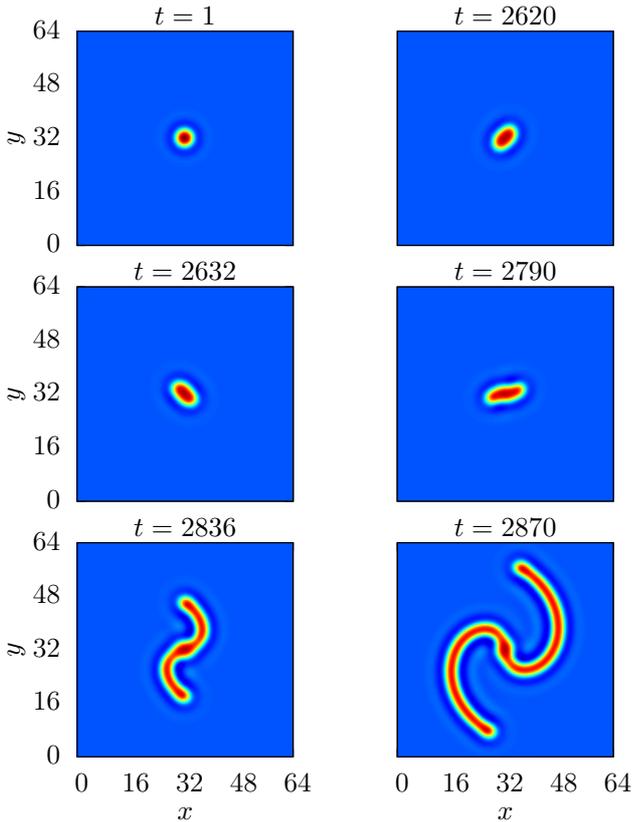}

 \caption{Formation of a spreading spiral observed in direct numerical simulations in two dimensions. Parameters are: $\alpha=0.31$, $\tau=5.7$, $L_x=L_y=64.0$, $a_1=2.0$, $a_2=\frac{4}{3}$, $Y_0=-0.4$, $C=1.0$, $A=2.0$, $B=4.0$.}
 \label{fig:spirale}
\end{figure}
\end{section}
However, in the following section, the focus lies on the description of depinning localized structures, i.e. the parameter regime of small amplitudes of the inhomogeneity $A$, where the translational mode is the first to become destabilized by time-delayed feedback.
\begin{section}{Semi-Analytic Description of a Pinned Localized Structure}\label{sec:analytic}
In this section we are going to focus on the behavior of a localized structure pinned on the inhomogeneity and its destabilization, leading to an oscillating or drifting structure (see Fig. \ref{fig:1d1szeit}). We discuss two different approaches to describe the system in the vicinity of the first bifurcation at $\alpha_\text{crit}$. For the sake of simplicity, we restrict the analysis to a one-dimensional system. However both approaches can be easily generalized to more than one dimension and may also be applied to other inhomogeneous systems.

\begin{subsection}{Approach 1: Order-Parameter Equation}\label{sec:OPE}
As a first approach we use an ansatz similar to \cite{gurevich2014time}, that describes the solution of the system in the vicinity of the Hopf bifurcation as the stationary solution $q_0(\textbf x)$ with an additional perturbation $\tilde{q}(\textbf x,t)$ in the form of an oscillation in the spatial form of the translational eigenmode $\varphi(\textbf x)$:
\begin{align}
 q(\textbf x,t)&= q_0(\textbf x)+\tilde{q}(\textbf x,t) \\&=q_0(\textbf x)+\xi(t)\varphi(\textbf x)\text e^{\text i\omega t}+\overline\xi(t)\varphi(\textbf x)\text e^{-\text i\omega t}+{\xi}_0(\textbf x,t),\label{Ansatz1}
\end{align}
where $\xi(t)$ is a slowly varying complex order parameter, $\overline\xi(t)$ is its complex conjugate, $\omega=\text{Im}(\lambda_0)$ is the frequency of the oscillation at the bifurcation point and ${\xi}_0(\textbf x,t)$ accounts for further contributions of stable eigenmodes.

Inserting the ansatz \eqref{Ansatz1} into Eq. \eqref{DSH} and collecting only the terms of $\mathcal{O}\left(\text e^{\text i\omega t}\right)$ leads to:
\begin{align}
 \dot\xi\varphi+\text i \omega\xi\varphi&=\xi\mathfrak L'\varphi+\xi\mathfrak L''{\xi}_0 \varphi+\frac{1}{2}|\xi|^2\xi\mathfrak L'''\varphi\varphi\varphi\\&+\alpha\left(\xi(t)-\xi(t-\tau)e^{-\text i\omega \tau}\right)\varphi,\label{Orderomega}
\end{align}
where $\mathfrak L^{(n)}$ denotes the $n$th Fr\'echet derivative of the nonlinear righthandside of Eq. \eqref{DSH} without the delayed terms. Further, to estimate ${\xi}_0(\textbf x,t)$, we insert the ansatz \eqref{Ansatz1} into Eq. \eqref{DSH} and collect only the non-oscillating terms:
\begin{align}
 \dot{{\xi}}_0=\mathfrak L'{\xi}_0+|\xi|^2\mathfrak L''\varphi\varphi, \label{Order0}
\end{align}
 Note that the time-delayed feedback has been neglected in the last equation because ${\xi}_0(t)$ varies slowly in time compared to the order parameter $\xi(t)$. However, since $\xi_0(t)$ changes fast compared to $|\xi(t)|$ one can adiabatically eliminate the time evolution of ${\xi}_0$, which leads to:
\begin{align}
 \mathfrak L' X_0=-\mathfrak L''\varphi\varphi,\label{Randbedingungx0}
\end{align}
where ${\xi}_0= X_0 |\xi|^2$ has been introduced. The function $ X_0$ can be calculated by numerically inverting $\mathfrak L'$, i.e. by solving Eq. \eqref{Randbedingungx0}. Inserting the solution for ${\xi}_0$ and projecting on $\langle \varphi|$ finally yields the order parameter equation:
\begin{align}
 \dot\xi(t)=(\mu-\text i\omega)\xi(t)+b|\xi(t)|^2\xi(t)+\alpha\left(\xi(t)-\xi(t-\tau)e^{-\text i\omega \tau}\right)\,, \label{OPE1}
\end{align}
where $\mu$ is the eigenvalue of $\mathfrak L'$ corresponding to the eigenfunction $\varphi$ and 
$$
b=\frac{ \langle\varphi|\mathfrak L'' X_0\varphi\rangle}{\langle\varphi|\varphi\rangle}+\frac{1}{2}\frac{\langle\varphi|\mathfrak L'''\varphi\varphi\varphi\rangle}{\langle\varphi|\varphi\rangle}\,.
$$ 
The notation $\langle\cdot|\cdot\rangle$ stands for the scalar product defined by integration over the full domain.

Without time-delay, Eq. \eqref{OPE1} takes the normal form of a supercritical Hopf bifurcation below the bifurcation point, since $\mu,\,b<0$. Without the delay, as well as for sufficiently small values of $\alpha$, the stable solution of Eq. \eqref{OPE1} is $\xi=0$ corresponding to a stable solution $q(\textbf x,t)= {q}_0(\textbf x)$. For larger values of $\alpha$ the complex order parameter $\xi$ starts to oscillate between its real and imaginary part with a constant amplitude $|\xi|$. One can calculate the resulting maximum shift, i.e. the amplitude $\textbf R_\text{max}$ of the oscillation of the localized structure, by evaluating
\begin{align}
 q(\textbf x,t)={q}_0(\textbf x)+|\xi|\varphi(\textbf x)+ X_0 |\xi|^2 \label{shift}.
\end{align}
The shift of the maximum of $ q(\textbf x,t)$ compared to the maximum of ${q}_0(\textbf x)$ yields the amplitude of the oscillations $\textbf R_\text{max}$.

Solving Eq. \eqref{OPE1} numerically with a classical Runge-Kutta 4 timestep for different values of $\alpha$, one can compare the maximum shift resulting from Eq. \eqref{shift} with the amplitude of the oscillations from the direct numerical simulations. As can be seen in Fig. \ref{fig:OPE}, the order parameter model shows a bifurcation at $\alpha_\text{crit}$, i.e. the approximations made for the derivation of the order parameter equation seem to be justified in the direct vicinity of the bifurcation point. However, for larger values of $\alpha$, the model quickly looses its validity and a signficant difference between the direct numerical simulations and the order parameter model can be observed.
\begin{figure}[h]
 \centering
 \includegraphics[width=0.5\textwidth]{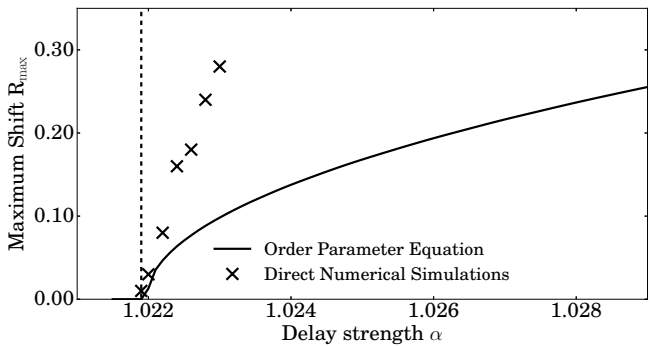}
 \caption{Maximum amplitude of the oscillations $\textbf R_\text{max}$ in dependence of the delay strength $\alpha$ obtained by direct numerical simulations (crosses) and by the order parameter model  \eqref{OPE1} (black solid line). The dotted line marks the first Hopf bifurcation point $\alpha_\text{{crit}}=1.0219$  found both in the direct numerical simulations and in the order parameter model.  However, Eq.~\ref{fig:OPE} is only valid in the direct vicinity of $\alpha_\text{{crit}}$ and looses its validity for increasing delay strengths.}
 \label{fig:OPE}
\end{figure}
\end{subsection}
\begin{subsection}{Approach 2: Overdamped Particle in a Potential Well}

The main idea of the second approach is to describe the oscillations occurring after the bifurcation at $\alpha_\text{crit}$ as the overdamped dynamics of a particle in a potential well generated by the inhomogeneity, where the time-delayed feedback acts as a driving force. Therefore we decompose the r.h.s. of Eq. \eqref{DSH} into a homogeneous part $N_\text{hom}$ containing everything but the delayed terms and the inhomogeneity, an inhomogeneous part $N_\text{inh}$ containing the inhomogeneity, and the time-delayed terms, i.e.:
\begin{align}
 \partial_t q(\textbf x,t)=N_\text{hom}[ q]+N_\text{inh}[\textbf x]+\alpha({q}(\textbf{x},t)-{q}(\textbf{x},t-\tau))\,.\label{DSHD}
\end{align}
We assume the solution $q(\textbf x,t)$ to be constant in shape, i.e., we neglect any shape deformations due to the oscillation. A similar ansatz without time-delay has been used in \cite{HERB2010pre}. It yields
\begin{align}
 q(\textbf x,t)=q_0(\textbf x-\textbf R(t))=q_{0\text h}(\textbf x-\textbf R(t))+w(\textbf x-\textbf R(t)), \label{Ansatz2}
\end{align}
where $q_0$ is the stationary solution of the inhomogeneous system, $q_{0\text h}$ ist the stationary solution of the homogeneous system, $ w(\mathbf{x}, t)$ is the shape deformation of the stationary solution caused by the inhomogeneity and $\textbf R(t)$ is the postition of the center of the localized structure. The goal is, to derive a differential equation that describes the time evolution of the position $\textbf R(t)$.

Inserting the ansatz \eqref{Ansatz2} into Eq. \eqref{DSHD} yields:
\begin{align}
 -\dot{\textbf R}(t)\partial_xq_0(\textbf x-\textbf R(t))=N_\text{hom}[\textbf q_{0}(\textbf x-\textbf R(t))]\nonumber\\+ N_\text{inh}[\textbf x]+\alpha\left({q}_0(\textbf{x}-\textbf R(t))-{q}_0(\textbf{x}-\textbf R(t-\tau))\right)\,.
\end{align}
In addition, expanding $N_\text{hom}[q_{0}(\textbf x-\textbf R(t))]$ around $q_{0\text h}(\textbf x-\textbf R(t))$ results in:
\begin{align}
 N_\text{hom}[q_{0}]=N_\text{hom}[q_{0\text h}]+\mathfrak L'[q_{0\text h}]w\nonumber\\+\frac{1}{2}\mathfrak L''[q_{0\text h}]w w+\frac{1}{6}\mathfrak L'''[ q_{0\text h}]www\,. \label{expansion}
\end{align}
Looking at Eq \eqref{expansion} one can easily verify, that $\langle \varphi_G(\textbf x-\textbf R(t))| N_\text{hom}[ q_{0}(\textbf x-\textbf R(t))]\rangle=0$, where $\varphi_G$ is the translational mode of the homogeneous system, i.e. a Goldstone mode. Indeed, $N_\text{hom}[q_{0\text h}]=0$, because $q_{0\text h}$ is a stationary solution of the homogeneous system. The linear term in $w$ vanishes, because $\mathfrak L'$ is a self-adjoint operator and the eigenvalue corresponding to $\varphi_G$ is $\mu=0$. The quadratic and cubic terms vanish, because even and odd functions are multiplied and integrated over the full domain. Projecting $\langle \varphi_G(\textbf x-\textbf R(t))|$ onto Eq. \eqref{DSHD} therefore leads to:
\begin{align}
  \dot{\textbf R}(t)=\frac{-1}{\langle \varphi_G(\textbf x)|\partial_x  q_0(\textbf x)\rangle}\Big(\langle \varphi_G(\textbf x)| N_\text{inh}[\textbf x+\textbf R(t)]\rangle \nonumber\\-\alpha \langle \varphi_G(\textbf x)| q_0 \big(\textbf x+\textbf R(t)-\textbf R(t-\tau)\big)\rangle \Big)\,.\label{OPE2}
\end{align}
The first term on the r.h.s. of Eq. \eqref{OPE2} yields a function 
$$
 F(\textbf R)=\frac{-\langle \varphi_G(\textbf x)|  N_\text{inh}[\textbf x+\textbf R(t)]\rangle}{\langle \varphi_G(\textbf x)|\partial_x q_0(\textbf x)\rangle}\,.
$$ 
that can be interpreted as the attracting force of the inhomogeneity acting on the localized structure. Figure \ref{fig:potential} depicts the potential well $ V(\textbf R)$ of the inhomogeneity, which is defined as: $-\partial_\textbf R V(\textbf R)= F(\textbf R)$.
\begin{figure}[h]
 \centering
 \includegraphics[width=0.5\textwidth]{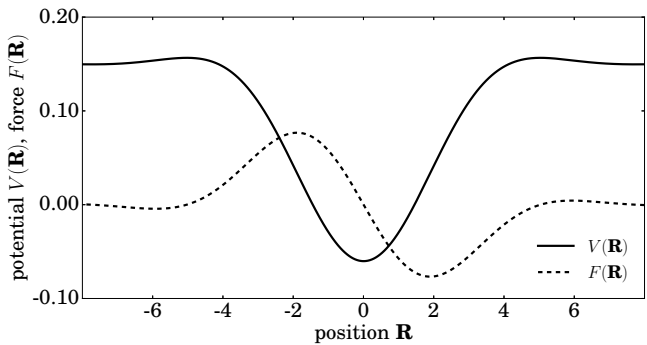}
  \caption{Potential well $V(\textbf R)$ and attracting force $F(\textbf R)$ of an inhomogeneity ${Y}$ calculated numerically for $A=0.2$, $B=4.0$.}
  %obtained by numerically evaluating the %integrals $\textbf F(\textbf R)=\frac{-\langle \boldsymbol\varphi_G(\textbf x)| \textbf N_\text{inh}[\textbf x+\textbf R(t)]\rangle}{\langle \boldsymbol\varphi_G(\textbf x)|\partial_x \textbf q_0(\textbf x)\rangle}$.}
 \label{fig:potential}
\end{figure}
Without time-delayed feedback, the potential $V(\textbf R)$ can also be used to estimate the basin of attraction of the inhomogeneity ${Y}$. Placing a localized structure in the direct vicinity of the inhomogeneity leads to the structure being pulled to the minimum of the potential at $\textbf R=\textbf 0$, since all integrals necessary for the calculation of $ F(\textbf R)$ vanish for $\textbf R=\textbf 0$. However, the potential also has two maxima in the periphery of the inhomogeneity, i.e. for larger values of $\textbf R$, the potential acts repelling on the localized structure. These results are in good agreement with the behavior of localized structures observed in direct numerical simulations described in section \ref{sec:Model}.

With time-delayed feedback ($\alpha\neq0$), the stable solution $\textbf R=\textbf 0$ gets destabilized for values of $\alpha>\alpha_\text{crit}$ leading to an oscillation in the potential well, where the time-delayed feedback acts as a driving force. Solving Eq. \eqref{OPE2} with a classical Runge-Kutta scheme yields the oscillatory dynamics of $\textbf R(t)$. Figure \ref{fig:potentialwell} shows the amplitude $\textbf R_\text{max}$ of the oscillations in comparison to the results from direct numerical simulations. As can be seen, the first bifurcation in the potential well model occurs at $\alpha_\text{{crit}}$, i.e. at the value expected from the linear stability analysis and direct numerical simulations, e.g. from evaluating Eq. \eqref{taukritisch}. The predictions from the potential well model \eqref{OPE2} are accurate throughout most of the parameter regime, where oscillations occur. Only close to the secondary instability at $\alpha_\text{{crit}2}$, where the localized structure depins from the inhomogeneity, notable differences 
between the potential well model and the direct numerical results can be observed. These differences can be ascribed to the deformation of the localized structure that is neglected in the presented potential well approach. The potential well model \eqref{OPE2} still reproduces the depinning, i.e. a process, where the localized structure escapes from the potential well due to a large driving force induced by time-delayed feedback. However, this secondary instability occurs at a value of $\alpha=1.079$, which is slightly larger than the value obtained from direct numerics $\alpha_\text{{crit}2}=1.072$.
\begin{figure}[h]
 \includegraphics[width=0.5\textwidth]{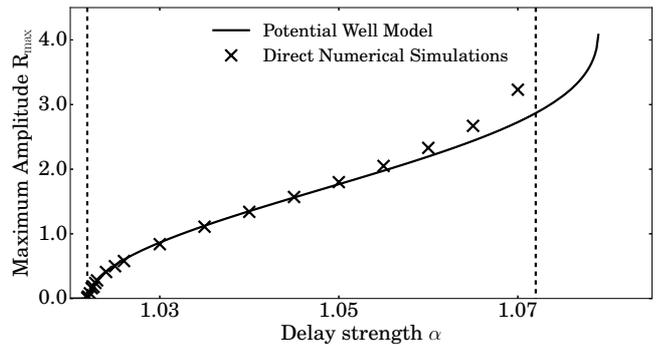}
 \caption{Maximum amplitude of the oscillations $\textbf R_\text{max}$ in dependence of the delay strength $\alpha$ obtained by direct numerical simulations (crosses) and by the potential well model  (black line). The dotted lines mark the two bifurcation points in the direct numerical simulations. The first bifurcation leading to an oscillation of the structure appears at the same value $\alpha_\text{{crit}}=1.0219$ in the direct numerical simulations and in the potential well model, respectively. The depinning instability occurrs in the potential well model at $\alpha>\alpha_\text{{crit}2}$, i.e. for a larger delay strength than in the direct numerical simulations.}
 \label{fig:potentialwell}
\end{figure}
\end{subsection}

\end{section}
\begin{section}{Conclusions}

An inhomogeneous Swift-Hohenberg model that describes pattern formation in the transverse plane of an optical cavity subjected to time-delay feedback is investigated in details. A linear stability analysis of the system has shown, that in the presence of spatial inhomogeneities, the discrete eigenvalues corresponding to localized eigenfunctions get altered, leading to different complex dynamical behaviors. In particular,  the eigenvalues of the translational eigenfunction change drastically and become complex. This behavior is attributed to the spatial inhomogeneity that breaks the translational symmetry of the system. The interplay between destabilized translational modes due to delay and attracting inhomogeneity leads to an oscillatory behavior in both transverse dimensions of the cavity. For larger delays (i.e. larger delay-time $\tau$ or larger delay-strength $\alpha$), the localized structure depins from the inhomogeneity. In the last part of the paper, we have presented two different approaches to treat the interplay between inhomogeneity and drift analytically:  (i) The derived order parameter equation presented in section \ref{sec:OPE} only reproduces the behavior of the system in the vicinity of the bifurcation point at $\alpha_\text{{crit}}$ and is less accurate throughout the rest of the parameter regime. One possible explanation for the small scope of the order parameter model is that the assumption that the frequency of the oscillation can be aproximated as the imaginary part of the eigenvalue $\omega=\text{Im}~\lambda$ is only valid in the  neighbourhood of the bifurcation point (cf. Fig. \ref{fig:omega}). However, due to the form of the order parameter equation \eqref{OPE1}, it provides a better understanding of the manifestation of the Hopf instability;  (ii) The potential well model which qualitatively describes the behavior of the localized structure in the complete parameter regime, including the first and the secondary instability. It is also quantitatively accurate for most values of $\alpha$ that are not too close to the secondary instability at $\alpha_\text{{crit}2}$. The potential well model has proven itself very useful in providing a simple and instructive way to deal with inhomogeneous systems, where the complex dynamics of a delay-driven localized structure in the vicinity of an inhomogeneity are reduced to the mechanical problem of an overdamped particle in a potential well with a driving force. However, it would be beneficial to refine the method by also considering shape deformations of the oscillating localized structure. 

The described dynamical solutions presented in this paper only depend on the competition between unstable translational modes and attracting inhomogeneities. Due to the generality of these results, we expect to observe similar dynamics in practical applications in nonlinear optical systems with inhomogeneities.

We thank Walter Tewes for fruitful discussions on the potential well approach and Markus Wilczek for helpful discussions of numerical issues.

\end{section}
\bibliographystyle{apsrev4-1}
\bibliography{Bibs.bib}
\end{document}